\documentclass[natbib]{svjour3}                     
\usepackage{graphicx}
 \usepackage{aps-bibstyle}  
\journalname{Space Science Reviews}
\def\FigureOne{{
\begin{figure}
\begin{center}
\includegraphics[width=\textwidth]{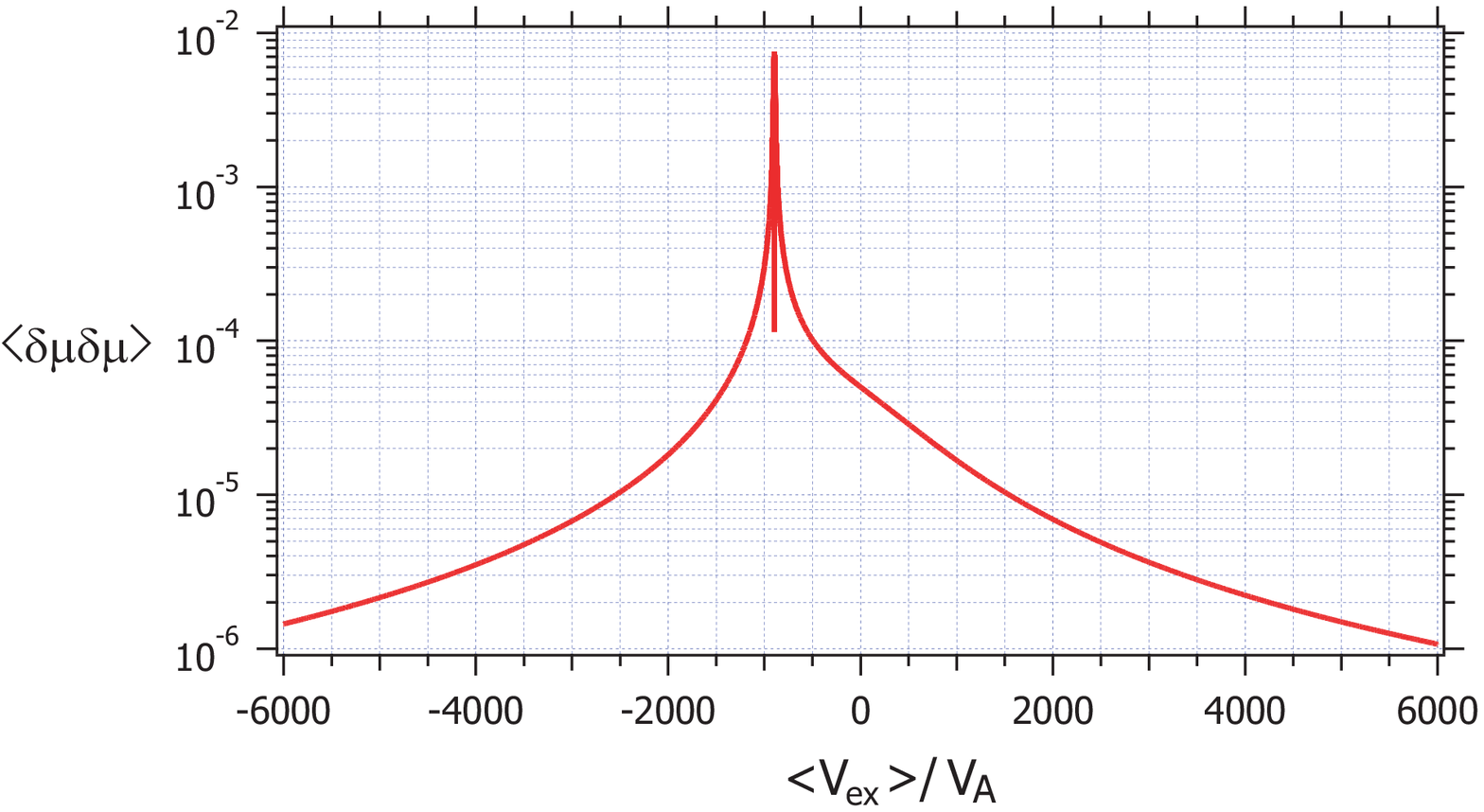}
\caption{
The result of test-particle calculation for
a parallel propagating wave with
$\omega =5 \Omega_{ci} = 2.72\times 10^{-3} \Omega_{ce}$, 
$k=k_x=2.044 \Omega_{ci}/V_A$, and $|{\vec B_w}|/B_0=10^{-4}$.
The variance of $\mu$, the cosine of the pitch angle,
is plotted against the average velocity $\langle V_{ex} \rangle$,
where the horizontal axis is normalized by the Alfv\'en velocity
$V_A$.
The peak at $\langle V_{ex} \rangle/V_A$ at $\sim -892.2$
shows the effect of the fundamental cyclotron resonance
condition.
}
\label{fig:1}
\end{center}
\end{figure}
}}

\def\FigureTwo{{
\begin{figure}
\begin{center}
\includegraphics[width=\textwidth]{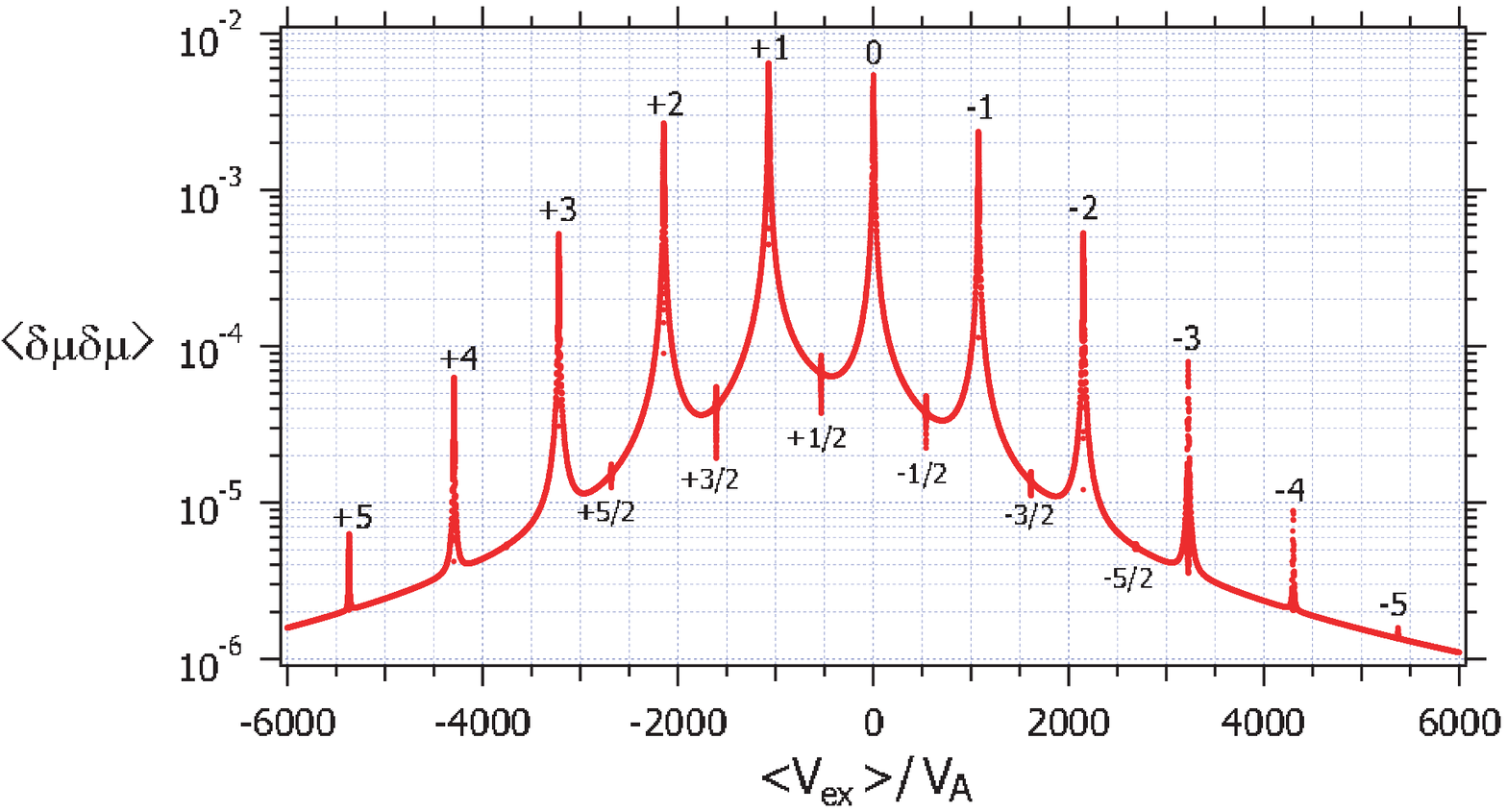}
\caption{
The same as Figure \ref{fig:1} except that
it is for a oblique propagating wave with
$\omega =5 \Omega_{ci} = 2.72\times 10^{-3} \Omega_{ce}$, 
$k=2.417 \Omega_{ci}/V_A$,
$k_x=1.709 \Omega_{ci}/V_A$,
$\theta = 45^{\rm o}$, 
and $|{\vec B_w}|/B_0=10^{-4}$.
Integers given above the peaks ($-5,-4, ..., ,5$) are
harmonic numbers in the resonance condition,
$\omega - k_x V_R = n \Omega_{ce}$.
Half integers given below the curve 
($-5/2, -3/2, ..., 5/2$)
are for the subharmonic resonance (see text).
}
\label{fig:2}
\end{center}
\end{figure}
}}

\def\FigureThree{{
\begin{figure}
\begin{center}
\includegraphics[width=0.5\textwidth]{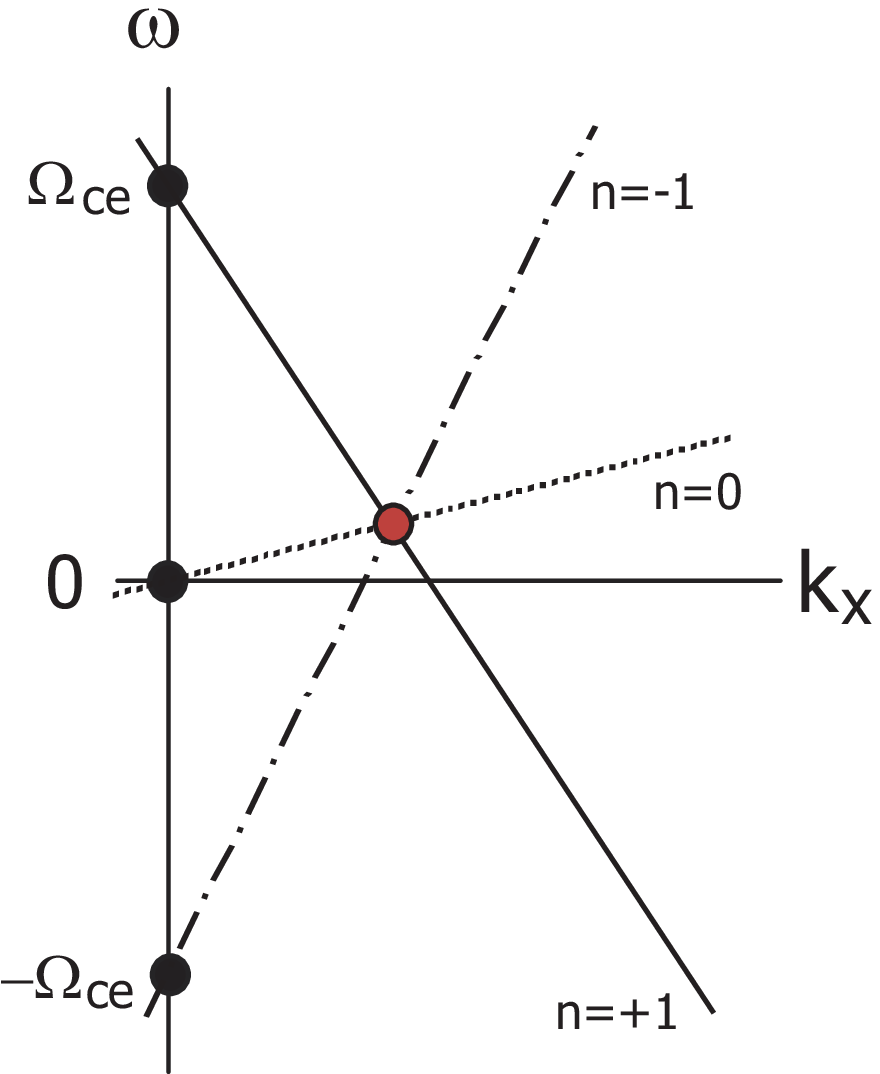}
\caption{
Schematic illustration of the three resonance conditions, 
$\omega - k_x V_R = n \Omega_{ce}$,
for ($n=0, \pm 1$) for an electron
with a whistler wave (red circle).
}
\label{fig:3}
\end{center}
\end{figure}
}}

\def\FigureFour{{
\begin{figure}
\begin{center}
\includegraphics[width=\textwidth]{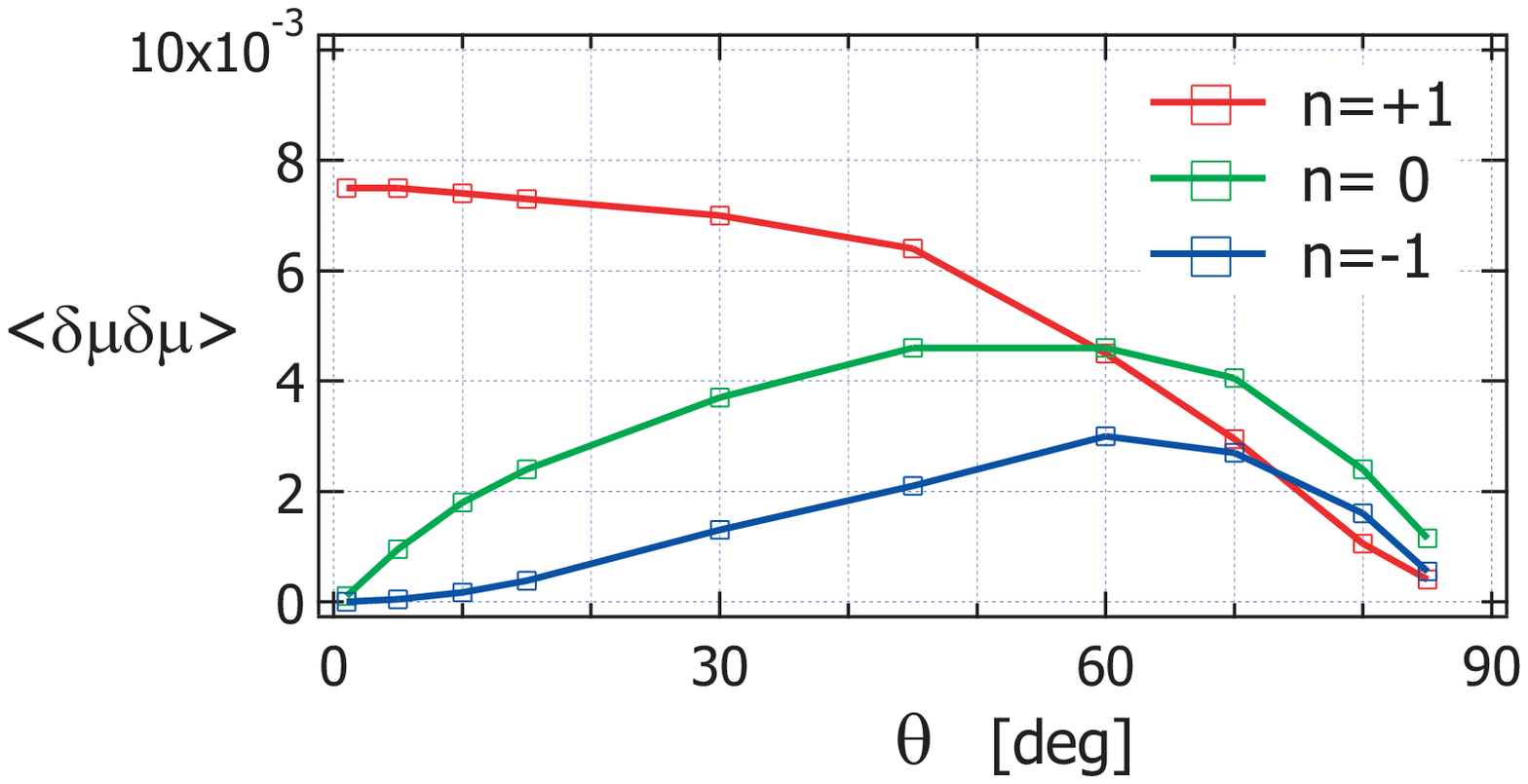}
\caption{
Angular dependence of $\langle \delta \mu \delta \mu \rangle$
for whistler waves at three different resonance conditions ($n=0, \pm 1$)
is shown for $\theta=$ $0^{\rm o} \sim 85^{\rm o}$.
The wave frequencies and relative amplitudes are fixed
at $\omega =5 \Omega_{ci} = 2.72\times 10^{-3} \Omega_{ce}$, 
and  $|{\vec B_w}|/B_0=10^{-4}$, respectively.
}
\label{fig:4}
\end{center}
\end{figure}
}}

\def\FigureAmanoHoshino{{
\begin{figure}
\begin{center}
\includegraphics[width=\textwidth]{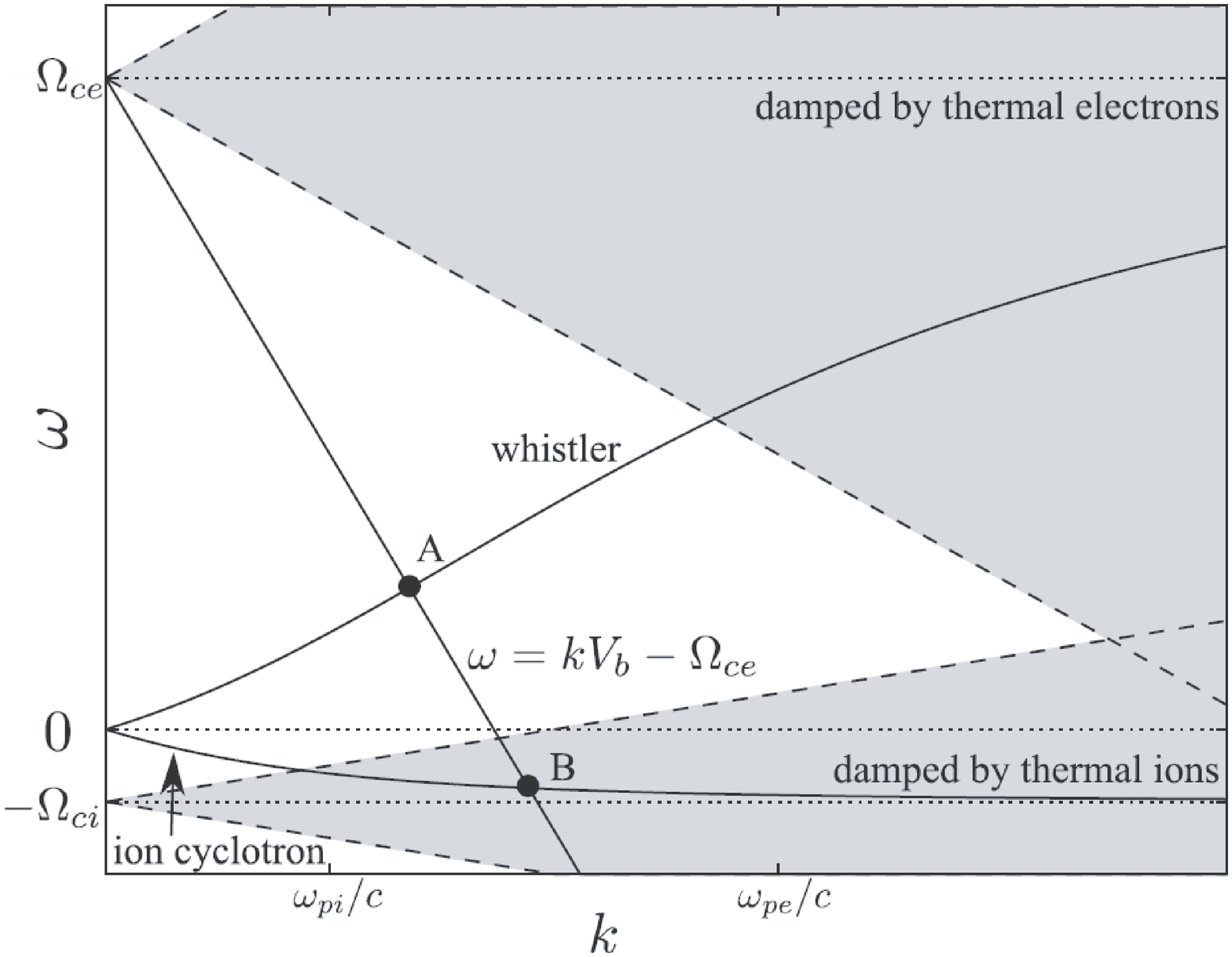}
\caption{From \cite{AmanoHoshino2010}.
Schematic dispersion diagram for whistler and ion-cyclotron waves
in an electron-ion plasma.
Positive (negative) frequency corresponds to the right-hand
(left-hand) polarization.
The cyclotron resonance condition (\ref{eqn:NRresonanceCondtion1st})
is also shown.
Waves in the shaded regions are strongly damped
by the cyclotron damping of thermal plasma particles
(electrons and ions).
}
\label{fig:AmanoHoshino}
\end{center}
\end{figure}
}}

\def\FigureOka{{
\begin{figure}
\begin{center}
\includegraphics[width=\textwidth]{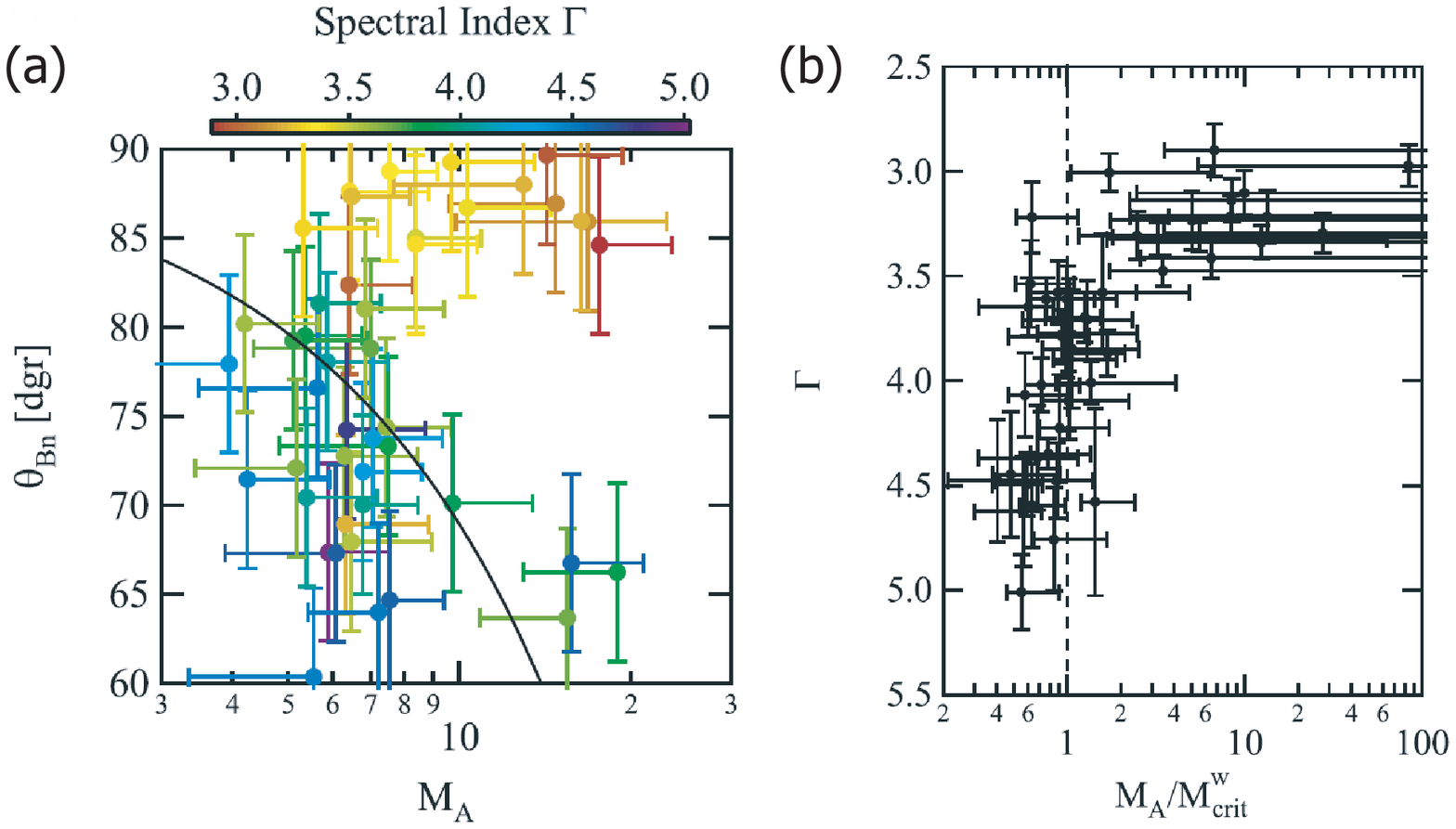}
\caption{From \cite{OkaETAL2006}.
(a) Observed spectral indices $\Gamma$ of suprathermal electrons
at the bow shock transition region are plotted onto the
two dimensional map of observed parameters ($M_A$, $\theta_{Bn}$).
$\Gamma$ is color-coded from orange (hard) to dark blue (soft).
A solid black curve shows the position of the
critical Mach number $M_{A,crit}$ where $\beta$ is set 1.
(b)  Observed spectral indices $\Gamma$ are plotted
against the Mach number normalized by $M_{A,crit}$
where observed $\theta_{Bn}$ is used to calculate $M_{A,crit}$.
}
\label{fig:Oka}
\end{center}
\end{figure}
}}

\def\FigureKuramitsu{{
\begin{figure}
\begin{center}
\includegraphics[width=\textwidth]{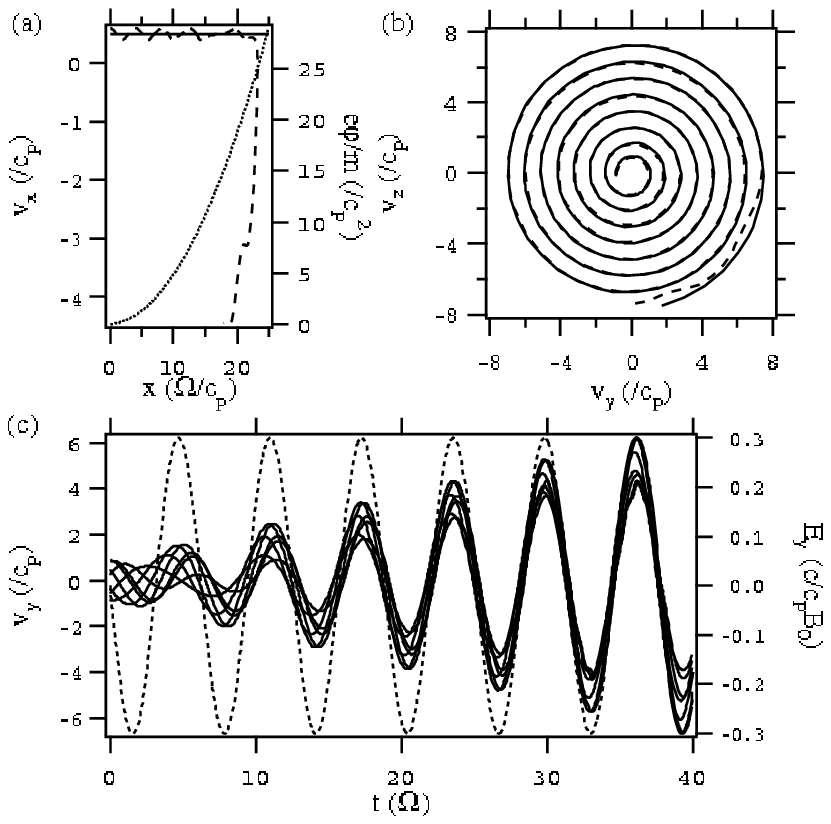}
\caption{Figures from \cite{KuramitsuKrasnoselskikh2005}.
Trajectories of ions satisfying a forced resonance condition 
with an electromagnetic wave under an external electrostatic 
field. Resonant (solid lines) and near-resonant (dashed lines) 
ions in (a) $v_x-x$ and (b) $v_z-v_y$ phase spaces. A dotted 
line in (a) denotes the spatial profile of the electrostatic 
field. (c) Time evolution of $v_y$ of the resonant ions 
(solid lines) and $E_y$ (dotted line).
}
\label{fig:Kuramitsu}
\end{center}
\end{figure}
}}

\def\FigureIsenbergVasquez{{
\begin{figure}
\begin{center}
\includegraphics[width=\textwidth]{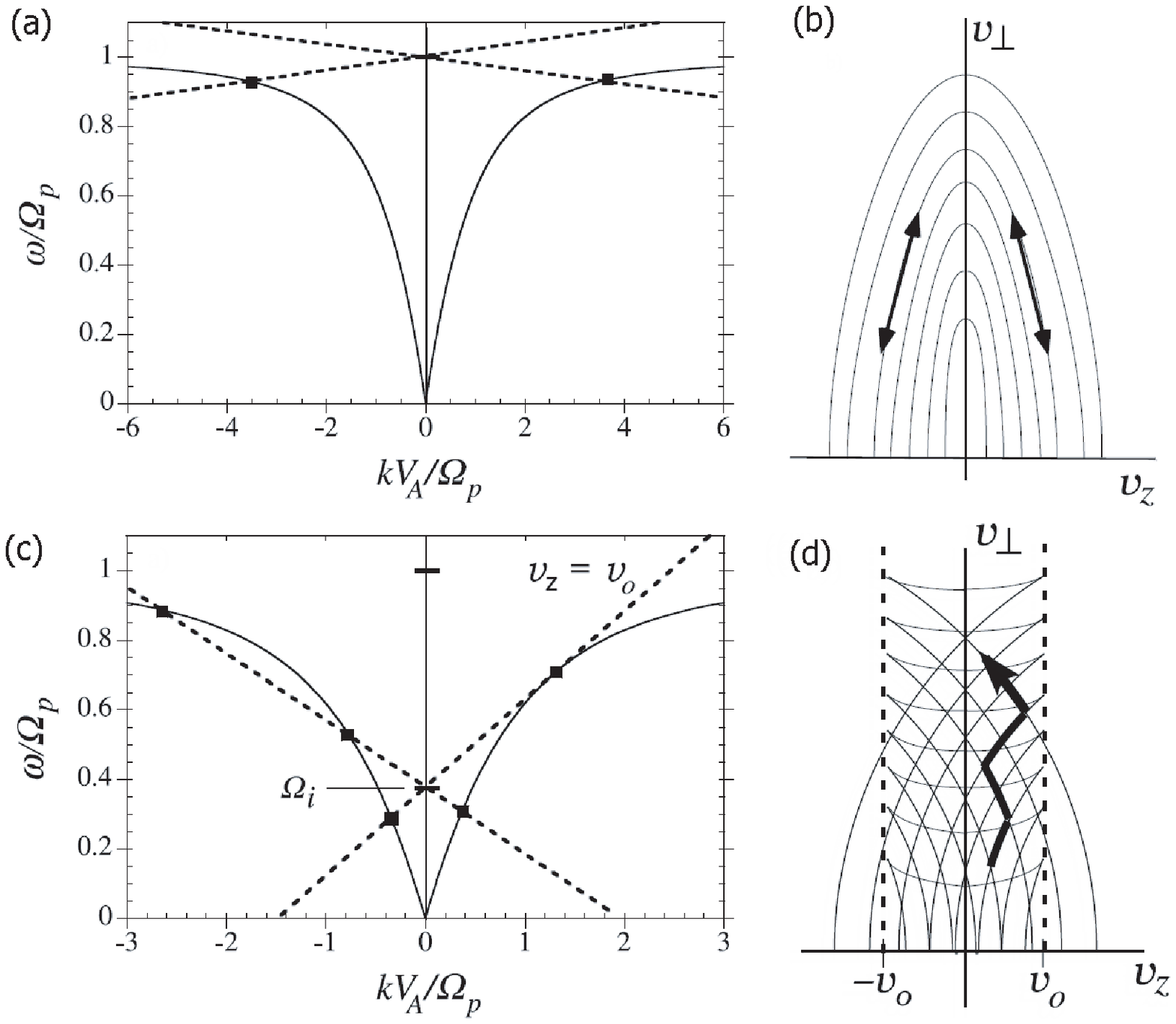}
\caption{From \cite{IsenbergVasquez2007}.
(a) The cyclotron resonance condition between protons
and parallel-propagating ion cyclotron waves.
The solid lines plot the wave dispersion relation,
while the dashed lines show the resonance condition for 
two particular proton parallel velocities with solid squares
indicating the resonance points with the waves.
(b) Resonant surfaces for protons in velocity space,
$
 ( v_z - \omega/k )^2 + v_\perp^2 = {\rm const.}
$
The cyclotron interaction causes protons to diffuse
along these surfaces, but not across them.
(c) and (d) Similar to (a) and (b) but for a minor ion species with $\Omega_i < \Omega_p$.
}
\label{fig:IsenbergVasquez}
\end{center}
\end{figure}
}}

\def\FigureMatsukiyoHada{{
\begin{figure}
\begin{center}
\includegraphics[width=0.65\textwidth]{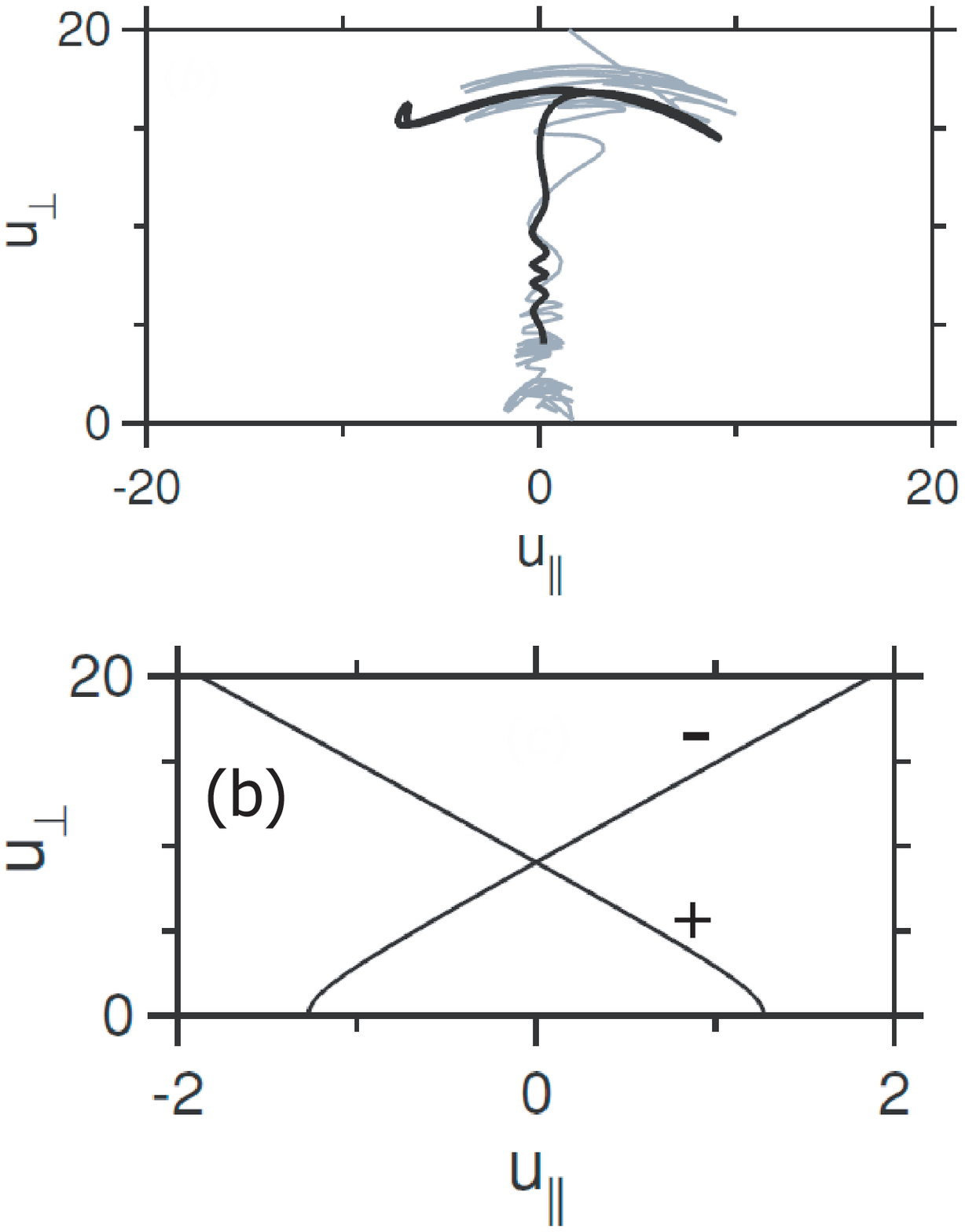}
\caption{Figures from \cite{MatsukiyoHada2009}.
(a) Two four-velocity trajectories (black and gray curves)
of relativistic electrons which are efficiently accelerated.
(b) Relativistic fundamental cyclotron resonance conditions 
(\ref{eqn:RELAresonanceCondtion2})
for positive (+) and negative (-) parallel wave numbers.
}
\label{fig:MatsukiyoHada}
\end{center}
\end{figure}
}}

\def\FigureOmura{{
\begin{figure}
\begin{center}
\includegraphics[width=0.60\textwidth]{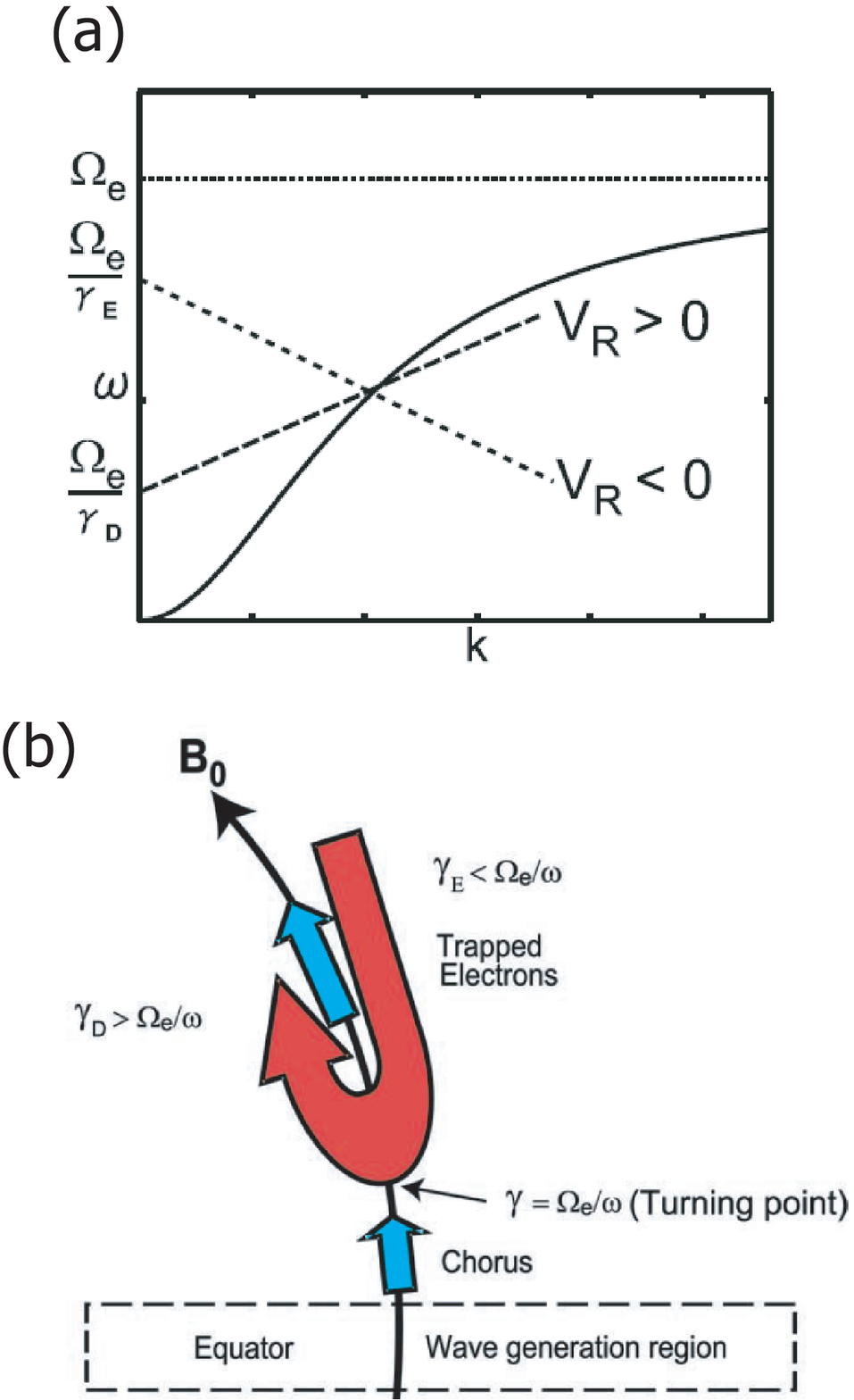}
\caption{Figures from \cite{OmuraETAL2007}.
(a) Fundamental relativistic cyclotron condition 
(\ref{eqn:RELAresonanceCondtion}).
(b) Illustration of `relativistic turning acceleration'.
Large amplitude chorus waves are assumed to be excited
near the equator and propagate from there to the
high latitude region along the field line.
An electron coming from the low altitude mirror point
toward the equator starts interacting with the chorus wave.
}
\label{fig:Omura}
\end{center}
\end{figure}
}}

\def\FigureKuramitsuTwo{{
\begin{figure}
\begin{center}
\includegraphics[width=0.7\textwidth]{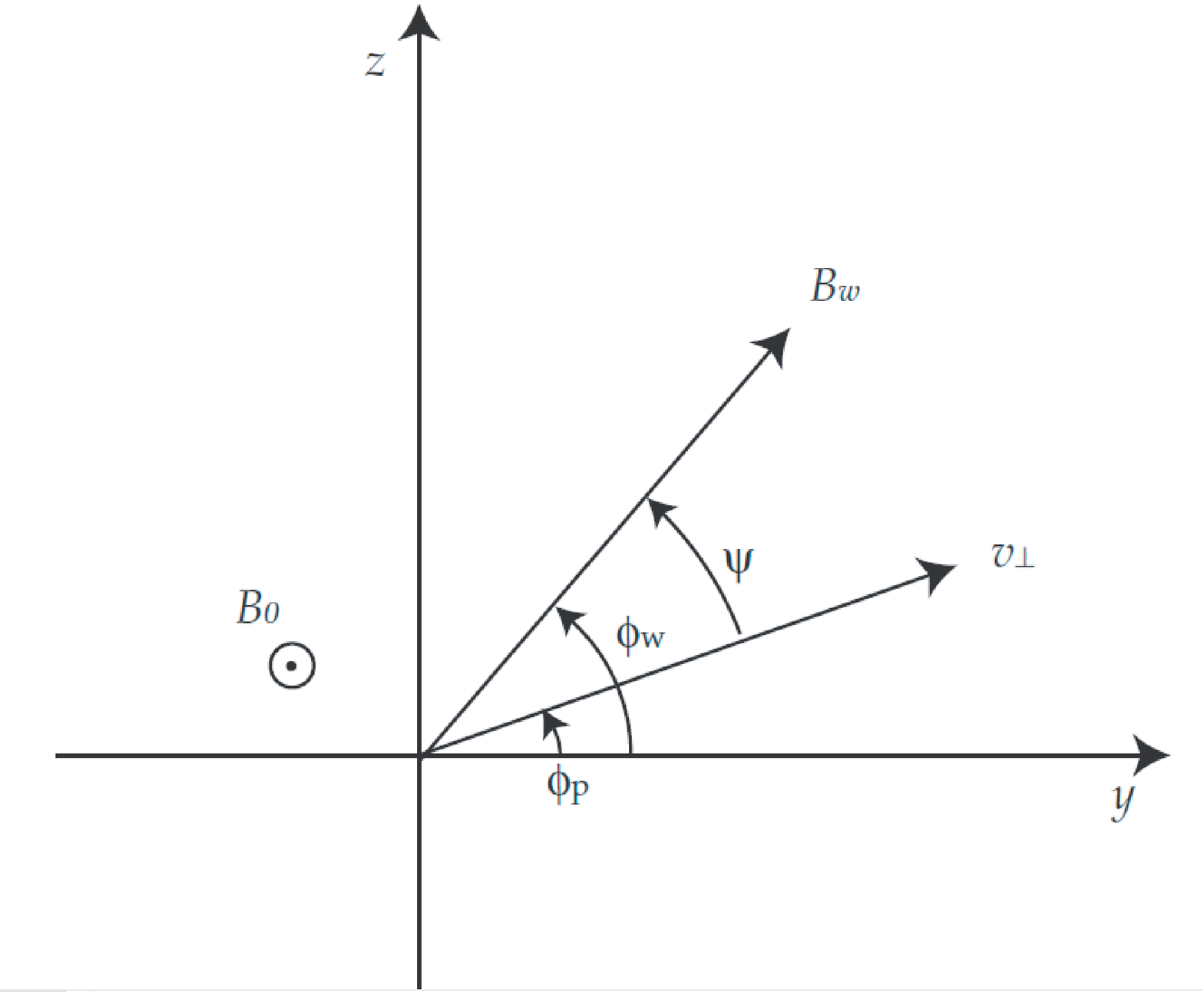}
\caption{Figure from \cite{KuramitsuKrasnoselskikh2005b}
showing the definition of angles, $\phi_p$, $\phi_w$, and $\psi$.
}
\label{fig:Kuramitsu2}
\end{center}
\end{figure}
}}

\def\FigureKuramitsuThree{{
\begin{figure}
\begin{center}
\includegraphics[width=0.95\textwidth]{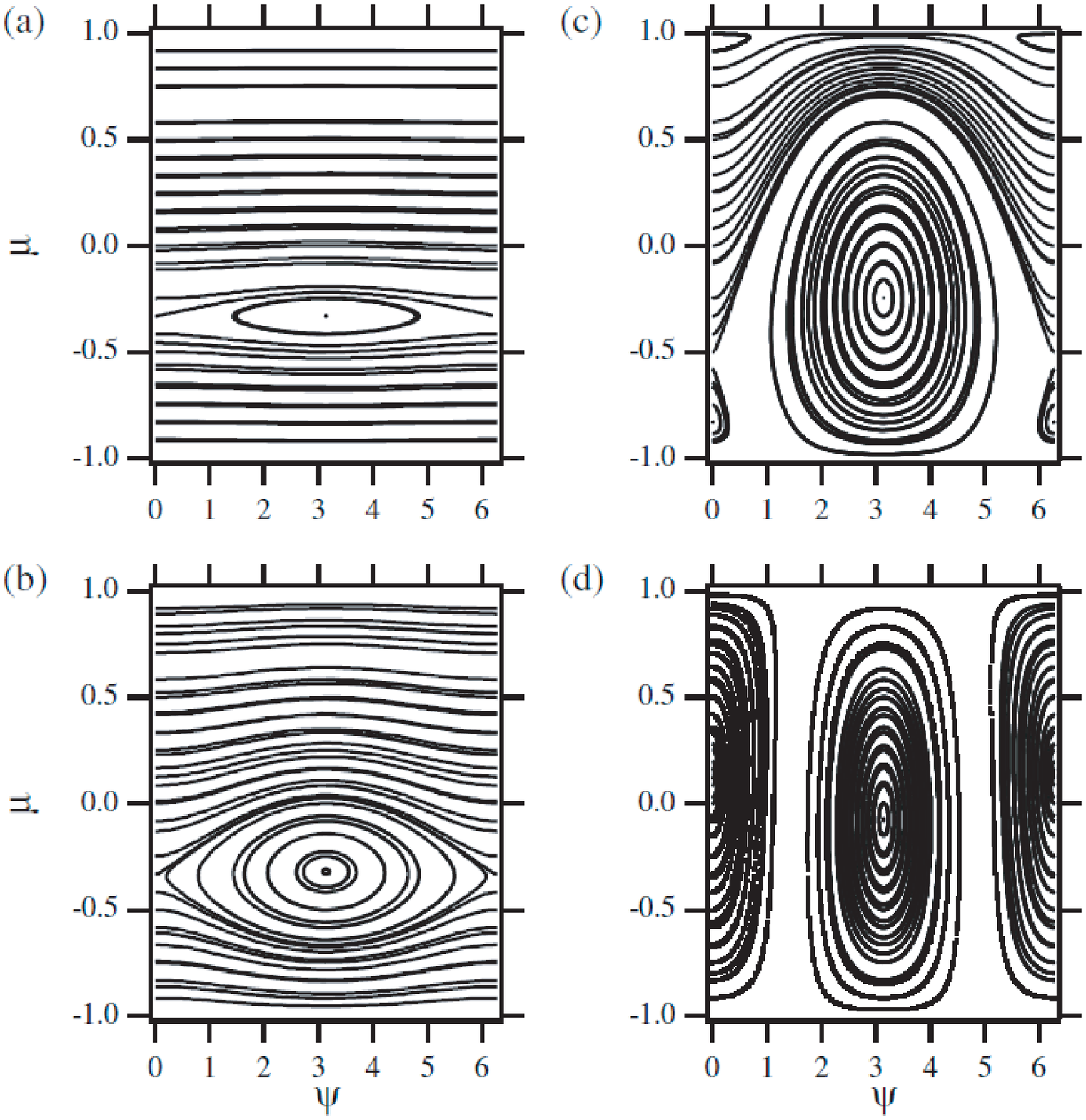}
\caption{Figure from \cite{KuramitsuKrasnoselskikh2005b}
showing particle trajectories in the $\mu-\psi$ phase space with $\kappa=3$
when (a) $b=0.01$, (b) $b=0.1$, (c) $b=1$, and (d) $b=10$.
One curve corresponds to one constant of motion $\chi$.
}
\label{fig:Kuramitsu3}
\end{center}
\end{figure}
}}

\begin{document}

\title{Cyclotron resonant interactions in cosmic particle accelerators
}

\titlerunning{Resonance in cosmic accelerators}

\author{Toshio Terasawa \and Shuichi Matsukiyo 
}

\authorrunning{Terasawa, Matsukiyo, and Hada}

\institute{T. Terasawa \at
   Institute for Cosmic Ray Research,
   University of Tokyo, 5-1-5 Kashiwa-no-ha, Kashiwa city, 
   Chiba 277-8582 Japan \\
   Tel.:+81-4-7136-5173\\
   Fax:+81-4-7136-3194\\
   \email{terasawa@icrr.u-tokyo.ac.jp} 
   \and
   S. Matsukiyo \at
   Earth System Science and Technology,
   Kyushu University, Kasuga city,
   Fukuoka 816-8580, Japan
}

\date{Received: 10 March 2012 / Accepted: 27 March 2012}

\maketitle

\begin{abstract}
A review is given for cyclotron resonant interactions in space plasmas.
After giving a simple formulation for the test particle approach,
illustrative examples for resonant interactions are given.
It is shown that for obliquely propagating whistler waves,
not only fundamental cyclotron resonance, but also other resonances,
such as transit-time resonance, anomalous cyclotron resonance,
higher-harmonic cyclotron resonance, and even subharmonic resonance
can come into play.
A few recent topics of cyclotron resonant interactions,
such as 
electron injection in shocks,
cyclotron resonant heating of solar wind heavy ions,
and relativistic modifications,
are also reviewed.


\end{abstract}

\section{Introduction}
\label{intro}
It is generally believed that
for efficient nonthermal particle acceleration
in various explosive cosmic events
strong plasma turbulence play the essential role.
For nonthermal particles to emerge from the pools
of majority thermal particle populations,
some selection mechanism(s) should work in turbulent environments.
The most likely selection mechanism is due to
resonance interaction
which assures the efficient energy/momentum transfer 
from the turbulence to nonthermal particles.
In this review, we will discuss
the cyclotron (and transit-time) resonance interactions
which would be viable
in various space plasma environments.

\section{Test particle approach}
\label{sec:TestParticleApproach}
In usual space-plasma environments, the electromagnetic fields 
have the background components (${\vec E_0}$, ${\vec B_0}$) and overlapping multi-wave components,
(${\vec E_w^\#}$, ${\vec B_w^\#}$) where $\#=1,2,...$ are the mode ID numbers.
The generalized Ohm's law gives the relation between ${\vec E_0}$ and ${\vec B_0}$
as 
$${\vec E_0} = -(1/c) {\vec V_{bulk}} \times {\vec B_0}, $$
where we write the bulk velocity of the plasma as ${\vec V_{bulk}}$
and neglect higher order terms caused by the spatial variations of
 ${\vec B_0}$
and electron thermal pressure.
If we make the coordinate transformation to the rest frame of the bulk plasma,
 ${\vec E_0}$ vanishes.
For a nonrelativistic electron, the equation of motion is\footnote{
Here we have chosen the electron motion.
Of course, we can equally take the ion motion
with a suitable modification of the choice of wave modes in the following consideration of resonant interactions.
The relativistic effect will be considered in 
{\S}\ref{sec:MatsukiyoHada} and {\S}\ref{sec:RelativisticTurningAcceleration}.}
\begin{equation}
\begin{array}{ccl}
\displaystyle{\frac{d \vec V_{e}}{dt}} + \Omega_{ce}  {\vec V_e} \times \frac{\vec B_0}{|\vec B_0|} &=&
\displaystyle{
-  \frac{e}{m_e }\sum_\# \left \{
     {\vec E_w^\#} + \frac{1}{c}  {\vec V_e} \times {\vec B_w^\#}
\right \}
} 
\end{array}
\label{eqn:EOMeV}
\end{equation}
\begin{equation}
\begin{array}{ccl}
\displaystyle{\frac{d \vec r_{e}}{dt}}   &~~~~~~~~~~~~~=~~~~~~~~~~~& {\vec V_{e}}~~~~~~~~~~~~~~~~~~~
\end{array}
\label{eqn:EOMeR}
\end{equation}
where $\Omega_{ce}$ is the electron cyclotron frequency $eB_0/m_e c$.
(\ref{eqn:EOMeV}) can be regarded as an oscillator equation of the eigen-frequency $\Omega_{ce}$
with the forcing term on the right hand side.
When we integrate (\ref{eqn:EOMeV}),
we evaluate
${\vec E_w^\#}$ and ${\vec B_w^\#}$ at
the electron position ${\vec r_e}(t)$ which is affected by the existence of the waves.
Therefore the system of equations (\ref{eqn:EOMeV}) and (\ref{eqn:EOMeR}) 
have nonlinearity, from which mode coupling effects among the wave components can appear.
However, if the amplitudes of waves are small enough, we can 
separately integrate (\ref{eqn:EOMeV}) for each wave,
\begin{equation}
\begin{array}{ccl}
\displaystyle{\frac{d \vec V_{e}}{dt}} +  \Omega_{ce}  {\vec V_e} \times \frac{\vec B_0}{|\vec B_0|}  &=&
\displaystyle{ 
-  \frac{e}{m_e } \left \{
     {\vec E_w^\#} + \frac{1}{c}  {\vec V_e} \times {\vec B_w^\#}
\right \} } 
\end{array}
\label{eqn:EOMeVforSingleWave}
\end{equation}
and then superpose the results afterward (the standard procedure in the quasi-linear treatment).
From here we choose a single wave mode and drop the superfix $\#$
\footnote{
Note, however, that there still remains self nonlinearity in (\ref{eqn:EOMeVforSingleWave})
for the mode \#,
where ${\vec E_w^\#}$ and ${\vec B_w^\#}$ are evaluated at the particle position ${\vec r_e}(t)$,
which is affected by ${\vec E_w^\#}$ and ${\vec B_w^\#}$ themselves.
This nonlinearity is the origin of subharmonic resonances to be discussed in {\S}\ref{sec:subharmonic}.
\label{note:selfnonlinearity}}.

For simplicity we consider electromagnetic waves in a cold uniform plasma
consisting of electrons and ions (protons) of equal density $n_0$.
We define the frequency and wave number of the waves as $\omega$ and $\vec k$, respectively.
If not otherwise noted, the $x$ axis is taken along the background magnetic field ${\vec B_0}$, 
and the $z$ axis is so chosen that the wave vector $\vec k$ is in the $x-z$ plane,
${\vec k} \equiv (k_x,~0,~k_z) = k(\cos \theta, ~0, \sin \theta)$.
Linear dispersion property is summarized in Appendix A.
We neglect the back reaction of electrons
toward the electromagnetic field
(test-particle approach).

\FigureOne

\section{Parallel propagating waves}
\label{sec:parallelPropagatingWave}
The first case is for a parallel propagating whistler wave
with $\omega =5 \Omega_{ci} = 2.72\times 10^{-3} \Omega_{ce}$, $k=k_x=2.044 \Omega_{ci}/V_A$,
and the relative wave amplitude $|{\vec B_w}|/B_0=10^{-4}$.
We inject $3.5 \times 10^4$ electrons with the initial conditions of 
\begin{equation}
r_e=(0,~0,~0)~{\rm and }~
  {\vec V_e}=(V_{_{ex,0}}, ~V_{e\perp} \cos \alpha,~V_{e\perp} \sin \alpha),
\label{eqn:Injection}
\end{equation}
where ($V_{_{ex,0}}, ~\alpha)$ are distributed uniformly in the two dimensional space, 
$(-6000 V_A,$ $~6000 V_A) \times (0, 2\pi)$, and $V_{e,\perp}$ is fixed at $1000 V_A$.
We integrate (\ref{eqn:EOMeR}) and (\ref{eqn:EOMeVforSingleWave}) for a long enough time $T$,
and then evaluate the averages of $V_{ex}$, $\mu$ (cosine of the pitch angle), and
$\delta \mu \delta \mu$ (the variance of $\mu$),
\begin{equation}
\begin{array}{lcl}
\langle V_{ex} \rangle &=& \displaystyle{ \frac{1}{T} \int_0^T V_{ex}(t) dt }\\ \\
\langle  \mu   \rangle &=& \displaystyle{ \frac{1}{T} \int_0^T \frac{V_{ex}(t)}{|{\vec V_e(t)}|} dt }\\ \\
\langle \delta \mu \delta \mu \rangle 
          &=& \displaystyle{ \frac{1}{T} \int_0^T \left (\frac{V_{ex}(t)}{|{\vec V_e(t)}|} -\langle \mu \rangle \right )^2 dt }
\end{array}
\end{equation}
In Figure \ref{fig:1} $\langle \delta \mu \delta \mu \rangle $ is plotted against $\langle V_{ex}\rangle$,
where we see a peak of $\langle \delta \mu \delta \mu\rangle $ at $\langle V_{ex}\rangle=-892.2 V_A \equiv V_R$,
where $V_R$ is the resonance velocity satisfying
\begin{equation}
\omega - k_x V_R = n \Omega_{ce}
\label{eqn:NRresonanceCondtion}
\end{equation}
with $n=+1$ (the fundamental cyclotron resonance condition).
This is the most natural resonance, in which
right-hand rotating electrons interact with
right-hand polarized whistler waves.
Historically-famous Kennel-Petscheck theory
(\cite{KennelPetschek1966})
for the loss-cone instability of whistler waves by
the trapped electrons in the radiation belt
mainly focused on this resonance.
Recently, \cite{AmanoHoshino2010} applied this resonance condition
to the electron injection process at shocks
({\S}\ref{sec:AmanoHoshino}).

Note that electrons in a monochromatic wave
are `trapped' around $V_R$
and their pitch angle diffusion does not occur.
In Figure 1 there is a dip at $\langle V_{ex}\rangle= V_R$, 
which is the manifestation of the trapping effect (Appendix B).
If, on the other hand, we superpose multi waves,
the interaction between the electron and one particular wave component
is limited within a finite coherent time $\tau$,
giving a contribution $\langle\delta \mu \delta \mu\rangle/\tau$
to the pitch-angle diffusion coefficient.
The task of quasilinear theory from this point of view
is to evaluate the coherent time $\tau$
for a given ensemble of wave components.
Since the literature on quasilinear theory is too vast,
we should limit the reference to only a few articles 
(e.g., 
\cite{KennelEngelmann1966,
      HallSturrock1967,
      HasselmannWibberenz1968,
      KulsrudFerrari1971,
      Jokipii1971,LeeLerche1974,Skilling1975,
      Schlickeiser1989a,Schlickeiser1989b,
      HollwegIsenberg2002,
      Schlickeiser2002,
      PetrosianLiu2004,
      Shalchi2009}).

\FigureTwo

\section{Oblique propagating waves}
\label{sec:obliquePropagatingWave}
Figure \ref{fig:2} is for an oblique propagating whistler wave
with $\omega =5 \Omega_{ci} = 2.72\times 10^{-3} \Omega_{ce}$, 
$k=2.417 \Omega_{ci}/V_A$,
$k_x=1.709 \Omega_{ci}/V_A$,
$\theta = \pi/4$, 
and the same relative wave amplitude $|{\vec B_w}|/B_0=10^{-4}$ as in {\S}\ref{sec:parallelPropagatingWave}.
We inject $5.0\times 10^4$ electrons with the same initial condition as
(\ref{eqn:Injection}).
Peaks labeled with integer numbers 
($n=-5,-4, ..., ,5$) correspond to the harmonic number $n$ in
the resonance condition (\ref{eqn:NRresonanceCondtion}).
It is noted that
there sometimes appear dips at the resonant velocities.
Again these dips are manifestation of
the trapping effect already seen in Figure \ref{fig:1}.
The resonance conditions for $n=0, \pm 1$ are
illustrated in Figure \ref{fig:3}.

\FigureThree

\subsection{Fundamental cyclotron resonance with $n=+1$}
The highest peak with $n=+1$ in Figure \ref{fig:2}
is for the fundamental cyclotron resonance condition,
\begin{equation}
\omega - k_x V_R = + \Omega_{ce}
\label{eqn:NRresonanceCondtion1st}
\end{equation}
which is already discussed in {\S}\ref{sec:parallelPropagatingWave}
for the purely parallel propagating case.

\subsection{Transit-time resonance with $n=0$}
The second highest peak in Figure \ref{fig:2}
is for the Landau resonance or transit-time resonance condition,
\begin{equation}
\omega - k_x V_R =0
\end{equation}
To see the origin of this resonance, we return to the $x$ component of the
equation of motion (\ref{eqn:EOMeVforSingleWave}),
\begin{equation}
\frac{dV_{ex}}{dt} = -\frac{e}{m_e} \left \{
E_{w,x} + \frac{1}{c} (V_{e,y} B_{w,z} - V_{e,z} B_{w,y} )
\right \}
\label{eqn:EOMeVx}
\end{equation}
where the second term in the parenthesis on the right hand side of (\ref{eqn:EOMeVx})
is the $({\vec V_e} \times {\vec B_w})_x$ term.
From the numerical inspection of (\ref{eqn:EOMeVx}),
we find that 
the relative contribution of the $E_{w,x}$ term is
less than 1\% of the $({\vec V_e} \times {\vec B_w})_x$ term.
(The dependence of $\langle \delta \mu \delta \mu \rangle$ on
$\langle V_{ex} \rangle$ is essentially unchanged,
if $E_{w,x}$ is artificially set zero.)
The $({\vec V} \times {\vec B})_x$ term
describes the mirror force exerted by
the variation of the magnetic field magnitude,
and this resonance is called `transit-time' resonance
(e.g., \cite{SchlickeiserMiller1998}).

\subsection{Anomalous cyclotron resonance with $n=-1$}
The resonance corresponding to the peak with $n=-1$
in Figure \ref{fig:2},
\begin{equation}
\omega - k_x V_R = - \Omega_{ce}
\label{eqn:NRresonanceCondtionAnomalous}
\end{equation}
is historically called `anomalous cyclotron resonance'
(or `anomalous Doppler effect': e.g., \cite{Ginzburg1960, Brice1964}).
(\ref{eqn:NRresonanceCondtionAnomalous}) is equivalent to
\begin{equation}
\frac{\omega}{k_x} - V_R = - \frac{\Omega_{ce}}{k_x}~~(<0)
\end{equation}
which means that
the wave is overtaken by the electron as $V_{R}$ 
exceeds the wave phase speed $\omega/k_x$ (here we set $k_x>0$).
Therefore, there is a reversal of the wave polarization
from the background plasma frame 
to the electron comoving frame.
The term `anomalous' stems from this reversal of polarization.
It is noted that an oblique propagating whistler wave is elliptically polarized
and consists of right-hand polarized (R) and left-hand polarized (L) subcomponents.
When the resonance condition (\ref{eqn:NRresonanceCondtionAnomalous}) applies, 
there are dual polarization reversals, namely
R and L subcomponents in the background plasma frame respectively become
left-hand and right-hand polarized in the electron comoving frame.
What the electron efficiently exchanges energy and momentum
at the $n=-1$ condition 
is the L subcomponent in the background plasma frame\footnote{
From this consideration it is naturally understood
why there is no anomalous resonance effect for a parallel propagating whistler wave
(Figure 1): An electron satisfying the $n=-1$ resonance condition
feels the whistler wave {\it left}-hand polarized in its own comoving frame
and does not exchange energy and momentum with the wave efficiently.
}.
\cite{Levinson1992} utilized this resonance condition to excite oblique whistler waves
by electron beams injected from the shock front.

\subsection{Cyclotron higher-harmonic resonance at $n$ with $|n| \ge 2$}
Cyclotron higher-harmonic resonance phenomena for 
oblique and perpendicular propagating waves 
are well known (see, e.g., Chapter 10 of \cite{Stix1992}).
Its application to Bernstein wave observed in the Jovian magnetosphere
is one of the classical examples (\cite{BarbosaKurth1980}).

\subsection{Cyclotron sub-harmonic resonance at half integers}
\label{sec:subharmonic}
In Figure \ref{fig:2}
there are some minor structures at subharmonic numbers, $$n=-5/2, -3/2, ..,3/2, 5/2,$$
where the dip structures (trapping effects) are seen to dominate over the small peak structures.
These are manifestations of cyclotron sub-harmonic resonance interaction,
which was first discussed by \cite{Smirnov1968}, 
and then followed by \cite{TerasawaNambu1989}
for perpendicular propagating magnetosonic waves.
There are renewed interests on this particular resonance
for oblique Alfv\'en waves 
(e.g., \cite{ChenEtAl2001, WhiteEtAl2002, LuChen2009}).
The origin of this resonance is the self nonlinearity
noted in the footnote$^{\ref{note:selfnonlinearity}}$ of {\S}\ref{sec:TestParticleApproach}.

\FigureFour

\subsection{Angular dependence of $\langle \delta \mu \delta \mu \rangle$}
\label{sec:AngularDependence}
As shown in Figure 2, there are many possible resonance modes
for oblique propagating waves.  It is of interest to see
how the relative importance of these mode depends on the propagation angle $\theta$.
Figure 4 shows the dependence of $\langle \delta \mu \delta \mu \rangle$ 
on the propagation angle $\theta$ of whistler waves 
at three different resonance conditions ($n=0, \pm 1$).
In the quasi-parallel propagation regime ($\theta <\sim 45^{\rm o}$),
the fundamental resonance of $n=+1$ dominates over others.
On the other hand, in the quasi-perpendicular regime ($\theta >\sim 45^{\rm o}$),
the transit-time resonance ($n=0$) and the anomalous cyclotron resonance ($n=-1$)
become progressively more important as $\theta \rightarrow 90^{\rm o}$,
and exceed the fundamental resonance above $\theta \sim 60^{\rm o}-70^{\rm o}$.

\section{Recent topics of resonant particle interactions}
\label{recenttopics}

\FigureAmanoHoshino

\subsection{A critical Mach number for electron injection in shocks}
\label{sec:AmanoHoshino}
It has been known that the electron component of galactic cosmic rays
is about 1\% of the proton component (e.g., \cite{Schlickeiser2002}).
While the origin of this percentage is not well understood,
it is generally interpreted as the indication of
smaller injection/acceleration efficiency for electrons than ions.
For the case of ions, a part of their suprathermal component
is reflected by shock fronts to form a beam.
The beam ions then excite MHD waves through the
ion-beam-cyclotron resonance condition 
and are subject to the pitch angle scattering process
(\cite{WinskeLeroy1984}), which is the first step to the
diffusive shock acceleration process for the ions.
If we apply the same scenario to electrons,
difficulty arises at the level of the wave excitation.
Figure \ref{fig:AmanoHoshino} illustrate this difficulty
(\cite{AmanoHoshino2010}):
An oblique solid line running from the upper-left 
is the fundamental cyclotron resonance condition
(\ref{eqn:NRresonanceCondtion1st})
for the electrons with the beam velocity, $V_{\rm e, beam}$,
\begin{equation}
\omega - k_x V_{\rm e, beam} =  \Omega_{ce}
\label{eqn:EBCI}
\end{equation}
The interaction point `A' between the beam electrons and the whistler wave
is not available for the wave excitation since
the beam electrons with purely parallel velocity {\it absorb} whistler wave energy at this point.
The interaction point `B' is for the anomalous cyclotron resonance condition
between the beam electrons and ion cyclotron wave (left-hand polarized in the plasma rest frame).
However, in the interplanetary and interstellar plasmas of typical temperature,
there are a plenty of thermal ions which damp the ion cyclotron wave
via fundamental cyclotron resonance interaction (shaded region around `B')
so that efficient pitch angle scattering of electrons is prohibited.

What \cite{AmanoHoshino2010} proposed is to invoke the loss cone distribution
of beam electrons, which is the natural consequence of
the electron reflection process at the shock front
(shock drift acceleration process) and can provide
the free energy to the whistler wave excitation
at the interaction point `A'
if the damping effect by thermal electrons
(shaded region around the $\omega = \Omega_{ce}$ line,
namely $-k_x v_{e,th} + \Omega_{ce} < \omega < k_x v_{e,th} + \Omega_{ce} $
with the thermal velocity $v_{e,th}$ of background electrons)
is avoided.
These considerations lead to the criterion for the
shock Alfv\'en Mach number $M_A$ for the efficient electron acceleration,
\begin{equation}
M_A \ge M_{A,crit} \equiv \frac{\cos \theta_{Bn}}{2} 
  \sqrt{\displaystyle{ \frac{m_i}{m_e} \beta_e}}
\end{equation}
which agrees with the observational criterion
for electron acceleration
obtained at the earth's bow shock,
where the hard spectral indices of suprathermal electrons
($\Gamma \le 3.5$) are observed
preferentially for high Mach number bow shock
satisfying $M_A > M_{A,crit}$  (Figure \ref{fig:Oka} from \cite{OkaETAL2006}).

\FigureOka


\FigureKuramitsu

\subsection{Gyroresonant surfing acceleration}
\label{sec:KuramitsuKrasnoselskikh}
\cite{KuramitsuKrasnoselskikh2005} proposed a new type of surfing 
acceleration of ions resonating with a left-hand circularly polarized wave 
through the fundamental cyclotron resonance condition\footnote{
In this and next subsections we consider the resonant interaction of
ions and set the sign of $\omega$ positive for the left-hand polarized waves.
} with $n = +1$,
\begin{equation}
\omega - k_x V_R = \Omega_{ci}
\label{eqn:IonCyclotronResonance}
\end{equation}
It is known that a particle trapped by a monochromatic circularly 
polarized wave oscillates around the resonant point in a phase 
space. During a trapped oscillation, a particle energy as well as a 
relative phase between the particle and the wave is bounded. 
However, if an external force is imposed to keep the relative 
phase constant, the particle can continue to gain energy from 
the wave electric field.

Such a forced resonance condition is realized, for instance, in 
the foreshock region of a quasi-parallel shock where an 
electrostatic potential acts as the external force. 
Figure \ref{fig:Kuramitsu} shows completely resonant 
(solid lines) and near resonant (dashed lines) ion trajectories 
in $v_x - x$ (a) and $v_z - v_y$ (b) phase spaces in a given 
electrostatic potential whose spatial profile is represented with 
a dotted line in (a). The two particles are accelerated 
monotonically in perpendicular direction while keeping the 
parallel velocities (almost) constant at $v_x \approx V_R (= 0.5)$. 
In their test particle simulation all the resonant particles 
are eventually accelerated regardless of their initial 
gyro phases due to the effect of phase synchronization which 
is shown in Figure \ref{fig:Kuramitsu} (c) in which time 
evolution of $v_y$ of the resonant particles having a variety 
of initial gyro phases (solid lines) and $E_y$ component of the 
wave electric field (dotted line) are plotted.

\FigureIsenbergVasquez

\subsection{Cyclotron resonant heating of solar wind heavy ions}
\label{sec:SolarWind}
It has been known that heavy ions in the fast solar wind have
faster flow velocity and hotter temperature (especially in
the direction perpendicular to ${\vec B_0}$) 
than the proton population, the main population of the solar wind.
\cite{IsenbergVasquez2007}
have proposed a second-order Fermi
acceleration mechanism for the preferential perpendicular
heating of heavy ions by counterstreaming left-hand circularly polarized waves.
In Figure \ref{fig:IsenbergVasquez},
the case for protons ((a) and (b)) is contrasted with that for heavy ions ((c) and (d)):
The second-order Fermi process works rather slowly for protons.
This is because
the signs of the resonance velocities for parallel and anti-parallel propagating
waves are different,
and the acceleration of protons by these waves does not occur simultaneously.
On the other hand, minor heavy ions whose cyclotron frequency
$\Omega_{i}$ is smaller than $\Omega_p$
have multiple resonant interaction points
if the condition
$|v_z| < v_0$ is satisfied (Figure \ref{fig:IsenbergVasquez} (c))\footnote{
The authors took the $z$ axis along ${\vec B_0}$.
The limiting velocity $v_0$ for the multiple interactions
corresponds to the condition tangential to
the wave dispersion relation (Figure \ref{fig:IsenbergVasquez}(c)).
}.
Therefore,  heavy ions can be accelerated by multi waves simultaneously
as indicated by the thick arrow in Figure \ref{fig:IsenbergVasquez}(d).

\FigureMatsukiyoHada

\subsection{Relativistic particle acceleration in developing turbulence}
\label{sec:MatsukiyoHada}
\cite{MatsukiyoHada2009} found a very efficient acceleration
of relativistic particles in developing Alfv\'en waves
in an electron-positron pair plasma,
where the wave number and frequency spectra of the waves are developing
through parametric decay instabilities.
Figure \ref{fig:MatsukiyoHada} (a) shows two examples of trajectories
of electrons which are efficiently accelerated
along the vertical line ($u_\parallel =0$).
The gray line denotes an accelerated electron in the course of 
the decay instability self-consistently reproduced in a 
one-dimensional full particle-in-cell simulation, while the 
black line is obtained from a test particle simulation in 
modeled electromagnetic fields.
This acceleration is interpreted in the following scenario:
In the developing turbulence, there exist waves 
propagating both directions parallel and anti-parallel to
the background magnetic field. 
The fundamental cyclotron resonance condition ($n=+1$)
for relativistic electrons is $V_x = V_R$
where $V_R$ satisfies
\begin{equation}
\omega - k_x V_R =  \frac{\Omega_{ce}}{\gamma} 
\label{eqn:RELAresonanceCondtion}
\end{equation}
With the four velocity
$(u_\parallel, u_\perp)$ $\equiv (v_x\gamma/c, ~v_\perp\gamma/c)$,
(\ref{eqn:RELAresonanceCondtion}) is rewritten as\footnote{
Note the following relation for the Lorentz factor,
$$
\gamma \equiv \{1-(v_x^2+v_\perp^2)/c^2 \}^{-1/2}
=(1+u_\parallel^2+ u_\perp^2)^{1/2}.
$$},
\begin{equation}
\left ( \frac{\omega}{\Omega_{ce}} \right ) 
\left ( 1+u_\parallel^2+ u_\perp^2 \right )^{1/2} 
  = 1 +\left ( \frac{k_xc }{\Omega_{ce} } \right ) u_\parallel 
\label{eqn:RELAresonanceCondtion2}
\end{equation}
Figure \ref{fig:MatsukiyoHada} (b)
shows the relation between $u_\parallel$ and $u_\perp$ given by (\ref{eqn:RELAresonanceCondtion2}),
where the frequency $\omega/\Omega_{ce}= 0.11 $ and
the parallel and anti-parallel wave numbers $k_x c /\Omega_{ce}= \pm 0.65 $
are chosen.
It is remarkable that at $u_\parallel=0$ the two resonance curves
overlap so that an electron can interact with two counterstreaming
waves at the same time and be efficiently accelerated.
This scenario is similar to the one for heavy ions (\S\ref{sec:SolarWind})
except that the former is made possible by the relativistic reduction of the cyclotron frequency
($\Omega_{ce}/\gamma < \Omega_{ce}$) instead of the intrinsic difference
in cyclotron frequencies for the latter ($\Omega_i < \Omega_p$).

\FigureOmura

\subsection{Relativistic turning acceleration}
\label{sec:RelativisticTurningAcceleration}
Recently, 
it is suggested that
relativistic electrons in the earth's radiation belt 
are coherently accelerated by large amplitude whistler waves
satisfying the relativistic cyclotron resonance condition
 (\ref{eqn:RELAresonanceCondtion})
(e.g., see \cite{OmuraETAL2007,FuruyaETAL2008,TaoBortnik2010}).
Large amplitude whistler waves are actually observed
as `chorus' emissions (\cite{CattellEtAl2008, CullyEtAl2008}).
The essence of the `turning acceleration' model
(\cite{OmuraETAL2007,FuruyaETAL2008}) is the relativistic reduction of the cyclotron frequency
in (\ref{eqn:RELAresonanceCondtion}).
Figure \ref{fig:Omura} shows two cases with $V_R<0$ (with $\gamma=\gamma_E$)
and $V_R>0$ (with $\gamma=\gamma_D>\gamma_E$).
An electron with $\gamma_E$ and $V_R<0$ 
is assumed to start interacting with the (counter-streaming) chorus wave.
Since the chorus wave has a large amplitude, the electron is trapped
and accelerated to increase $\gamma$ (i.e., reduce $V_R$)
coming to the turning point with $V_R =0$.
After the turning point, the electron moves toward
the same direction of the wave propagation,
so that the wave-particle interaction to continue long enough
to further accelerate the electron up to $\gamma = \gamma_D$,
where the electron is detrapped.


\section{Concluding remarks}
\label{conclusion}

We have given an overview
on the importance of resonant wave-particle
interaction processes in space plasmas.
Although each of elementary processes are well
known one by one, there is a necessity to 
unify these processes.
For example, as we see in {\S}\ref{sec:obliquePropagatingWave}
to consider interaction process between particles and 
oblique propagating waves 
we need to take into account of all resonance interactions
with harmonic numbers $n$=$0, \pm 1, \pm 2, ...$,
and sometimes even with subharmonic numbers $n$=$\pm 1/2, \pm 3/2, ...$.

In {\S}5.1-3 we see the various roles of cyclotron resonant condition
in the processes, shock injection, gyroresonant surfing, and
ion heating in the solar wind.
Of course, the considerations for the elementary resonance processes
alone do not suffice for comprehensive understandings of 
targeting physical processes:
Injected electrons at the shock front ({\S}\ref{sec:AmanoHoshino})
or surfing-accelerated ions at the shock front  ({\S}\ref{sec:KuramitsuKrasnoselskikh})
should be accelerated further in diffusive shock acceleration process, for example,
to acquire relativistic energy.
For the dynamics of heavy ions in the solar wind,
the second-order Fermi heating mechanism  ({\S}\ref{sec:SolarWind})
is incorporated into an inhomogeneous coronal hole model
 with
the effects of gravity, charge-separation electric field, and
mirroring force in the decreasing solar magnetic field 
(\cite{IsenbergVasquez2009}).

In {\S}\ref{sec:MatsukiyoHada} and
{\S}\ref{sec:RelativisticTurningAcceleration}
We further see 
that the relativistic resonance condition makes
the interaction process very rich.
Note that these considerations are for
the behavior of relativistic particles
in {\it nonrelativistic} turbulences.
While there are a few pioneering studies
in stochastic acceleration process
in {\it relativistic} turbulences
(e.g., \cite{ViratanenVainio2005,OSullivanETAL2009}),
further works on relativistic turbulence are
 needed to get comprehensive understanding, for example,
of the acceleration process of ultra-high-energy cosmic rays
(e.g., \cite{Hillas1984}).

Since we have limited the discussion to the waves
in cold plasmas,
wave modes raising from the finite temperature effect 
are outside of our scope.
Among these modes, kinetic Alfv\'en waves have
been a focus of extensive investigation.
Here we refer the interested readers to
a few representative papers
(e.g., \cite{HasegawaChen1976, LeamonETAL1999, LysakLotko1996}).

\begin{acknowledgements}
We would like to thank for valuable discussions and comments
to T. Hada, M. Scholer, M. A. Lee, and Y. Ohira.
The works by T.T. and S. M. get partial supports from the grants-in-aids
21540259 and 22740323 from Japan Society for the Promotion of Science,
respectively.
\end{acknowledgements}

\section*{Appendix A}

For the $j$ species ($j$=$i$ or $e$),
the plasma frequency is defined as $\omega_{pj} = \sqrt{4\pi n_0 e^2/m_j}$,
and the cyclotron frequency as $\Omega_{cj} = eB_0 /m_j c$.
The Alfv\'en velocity $V_A$ is defined as
$B_0/\sqrt{4\pi n_0 (m_i+m_e)}$,
which is chosen to be $10^{-4}c$ in the test particle calculation.
The ion mass $m_i$ is taken 1836 times $m_e$.

The wave electric field ${\vec E_w} \propto \exp i ({\vec k} {\vec r} - \omega t)$ should satisfy the following matrix relation
(see, e.g., Chapter 2 of \cite{Stix1992}),
\begin{equation}
 {\tilde M} {\vec E_w}=0,
\label{eqn:PolarizationProperty}
\end{equation}
with a matrix ${\tilde M}$ given as
\begin{equation}
{\tilde M} \equiv
\left \{
\begin{array}{ccc}
P-N^2 \sin^2 \theta & 0     & N^2 \sin \theta \cos \theta \\
0                   & S-N^2 & -iD                         \\
 N^2 \sin \theta \cos \theta & iD & S-N^2 \cos^2 \theta
\end{array}
\right \}
\end{equation}
where $N \equiv k c/\omega$ is a refractive index, and $P$, $S$, and $D$ are defined as
\begin{equation}
\begin{array}{c}
\displaystyle{
P=P(\omega)\equiv 1 - \frac{\omega_{pi}^2}{\omega^2} - \frac{\omega_{pe}^2}{\omega^2}} \\ \\
\displaystyle{
S=\frac{1}{2} (R(\omega)+L(\omega))  } \\ \\
\displaystyle{
D=\frac{1}{2} (R(\omega)-L(\omega))  } \\ \\
\displaystyle{
R(\omega)\equiv 1 - \frac{\omega_{pi}^2}{\omega (\omega+\Omega_{ci})} -  \frac{\omega_{pe}^2}{\omega (\omega-\Omega_{ce})}} \\ \\
\displaystyle{
L(\omega)\equiv 1 - \frac{\omega_{pi}^2}{\omega (\omega-\Omega_{ci})} -  \frac{\omega_{pe}^2}{\omega (\omega+\Omega_{ce})}}
\end{array}
\end{equation}
Note that the condition ${\rm det} {\tilde M} =0$ gives the wave dispersion relation.

From the first and second column of the matrix equation (\ref{eqn:PolarizationProperty}),
we have
\begin{equation}
\begin{array}{c}
E_{w,x} = \displaystyle{- \frac{N^2 \sin \theta \cos \theta}{P-N^2 \sin^2 \theta} ~E_{w,z} } \\ \\
E_{w,y} = \displaystyle{  \frac{iD                         }{S-N^2              } ~E_{w,z} }
\end{array}
\label{eqn:ExyzRelation}
\end{equation}
Now switching from the complex formulation to the real formulation,
we write the $xyz$ components of the electric field, ($E_{w,x}$, $E_{w,y}$, $E_{w,z}$), as
\begin{equation}
(E_{w,x}, E_{w,y}, E_{w,z})=
(g_x \cos \varphi, g_y \sin \varphi, g_z \cos \varphi)
\end{equation}
where the relative amplitudes of ($g_x$, $g_y$, $g_z$) are 
determined by (\ref{eqn:ExyzRelation}),
and $\varphi$ is the phase angle,
\begin{equation}
\varphi \equiv {\vec k} \cdot {\vec r} - \omega t + \varphi_0
\end{equation}
where $\varphi_0$ is the initial phase angle, respectively.
From the Faraday's law, we have
\begin{equation}
{\vec B_w} = N (\cos \theta, ~0, ~\sin \theta) \times {\vec E_w}
\end{equation}
from which the $xyz$ components of the magnetic field, 
($B_{w,x}$, $B_{w,y}$, $B_{w,z}$), are obtained as
\begin{equation}
\begin{array}{cllll}
B_{w,x} &=-& N  \sin \theta E_{w,y}&=-&N g_y \sin \theta \sin \varphi,\\ \\
B_{w,y} &= & N (\sin \theta E_{w,x} - \cos \theta E_{w,z})
                                   &= &N(g_x \sin \theta - g_z \cos \theta) \cos \varphi,\\ \\
B_{w,z} &= & N  \cos \theta E_{w,y}&= &N g_y \cos \theta \sin \varphi
\end{array}
\end{equation}

\FigureKuramitsuTwo

\FigureKuramitsuThree

\section*{Appendix B}
In a finite amplitude monochromatic electromagnetic wave
propagating parallel to the background magnetic field ${\vec B_0}$
charged particles can be phase-space trapped around
the resonant velocity
(see e.g.,
\cite{PalmadessoSchmidt1971,Helliwell1974,
Schmitt1976,Karpman1974,Matsumoto1979,
HoshinoTerasawa1985,KuramitsuKrasnoselskikh2005b}).
Here we follow the description in a recent article
by \cite{KuramitsuKrasnoselskikh2005b}.
Firstly, the gyrophase angle for an ion is defined as 
$\phi_p \equiv \tan^{-1} (v_z/v_y)$, 
the wave phase angle as $\phi_w \equiv kx - \omega t + \alpha_w$ ($\alpha_w$: the initial value),
and their difference as $\psi \equiv \phi_w-\phi_p$ 
(Figure \ref{fig:Kuramitsu2}).
Next, with the velocity of ions
in the wave rest frame $(u_x, u_\perp) = (v_x-\omega/k, v_\perp)$
the cosine of the pitch angle is defined as
$\mu \equiv u_x/|u|$,
where $|u| = (u_x^2 + u_\perp^2)^{1/2}$ is the constant of motion
(namely, the ion energy is conserved in the wave rest frame).
The second integral of the ion motion $\chi$ can be written as,
\begin{equation}
\chi = \frac{\kappa}{2} \left (
 \mu + \frac{1}{\kappa}
\right )^2
+ b (1-\mu^2)^{1/2} \cos \psi
\end{equation}
where 
$\kappa=k u/\Omega_{ci}$, and
$b$ the wave amplitude normalized by $|{\vec B_0}|$.
With $\chi$, the equation of motion is simplified as
\begin{equation}
{\dot \mu}  = -\frac{\partial \chi}{\partial \psi}, ~~~~~~~~
{\dot \psi} =  \frac{\partial \chi}{\partial \mu}
\end{equation}
Figure \ref{fig:Kuramitsu3} shows the ion trajectories
in the  $\mu-\psi$ phase space with $\kappa=3$
and $b=0.01$, 0.1, 1, 10.
When the wave amplitude is small ((a): $b=0.01$),
ions are trapped around 
$(\mu,\psi)=(-1/\kappa, \pi)$,
which is due to the linear cyclotron resonant interaction.
As the wave amplitude $b$ becomes larger ((c) and (d)),
the trapping region also becomes larger
and a new trapping region is born in a location
far from the original point $(-1/\kappa, \pi)$.


\end{document}